\documentclass[reprint, amsmath,amssymb, aps, superscriptaddress, floatfix]{revtex4-1}
\usepackage[utf8]{inputenc}
\usepackage{graphicx}% Include figure files
\usepackage{dcolumn}% Align table columns on decimal point
\usepackage{bm}% bold math
\usepackage{braket}
\usepackage{color}
\usepackage{units}
\usepackage{dsfont}%https://www.overleaf.com/project/5bf43b8a7f74cc6a37092aa5
\usepackage{float}
\usepackage{hyperref}% add hypertext capabilities
\usepackage{xcolor}
\begin{document}
\title{Quantum communication with ultrafast time-bin qubits}

% Force line breaks with \\
\author{Fr\'ed\'eric Bouchard}
\email{frederic.bouchard@nrc-cnrc.gc.ca}
\affiliation{National Research Council of Canada, 100 Sussex Drive, Ottawa, Ontario K1A 0R6, Canada}
\author{Duncan England}
\affiliation{National Research Council of Canada, 100 Sussex Drive, Ottawa, Ontario K1A 0R6, Canada}
\author{Philip J. Bustard}
\affiliation{National Research Council of Canada, 100 Sussex Drive, Ottawa, Ontario K1A 0R6, Canada}
\author{Khabat Heshami}
\affiliation{National Research Council of Canada, 100 Sussex Drive, Ottawa, Ontario K1A 0R6, Canada}
\affiliation{Department of Physics, University of Ottawa, Advanced Research Complex, 25 Templeton Street, Ottawa ON Canada, K1N 6N5}
\author{Benjamin Sussman}
\affiliation{National Research Council of Canada, 100 Sussex Drive, Ottawa, Ontario K1A 0R6, Canada}
\affiliation{Department of Physics, University of Ottawa, Advanced Research Complex, 25 Templeton Street, Ottawa ON Canada, K1N 6N5}

\begin{abstract}
The photonic temporal degree of freedom is one of the most promising platforms for quantum communication over fiber networks and free-space channels. In particular, time-bin states of photons are robust to environmental disturbances, support high-rate communication, and can be used in high-dimensional schemes. However, the detection of photonic time-bin states remains a challenging task, particularly for the case of photons that are in a superposition of different time-bins. Here, we experimentally demonstrate the feasibility of picosecond time-bin states of light, known as ultrafast time-bins, for applications in quantum communications. With the ability to measure time-bin superpositions with excellent phase stability, we enable the use of temporal states in efficient quantum key distribution protocols such as the BB84 protocol.
\end{abstract}

\maketitle

\section{Introduction}

Quantum communication is the branch of quantum technologies that deals with the distribution of quantum states of light to achieve a specific communication task. The most well known and developed quantum communication protocol is quantum key distribution (QKD) which allows two remote parties to share a secret key, enabling private communication~\cite{scarani:09,xu2020secure}. In its simplest form, QKD requires two mutually unbiased two-dimensional bases in which states can be prepared by a sender, Alice, transmitted, and measured by a receiver, Bob. High fidelity between the prepared state and the measured state is required. Loss and decoherence are therefore major challenges in practical QKD implementations, because they reduce, and eventually eliminate, the useful capacity of a communication channel. The decoherence encountered is strongly influenced by the quantum states employed and the mode of transmission. Various optical states have been exploited for QKD, including position bins, spatial modes, time bins, and polarization states~\cite{ding2017high,mirhosseini:15,bouchard2018experimental,mower2013high,inoue2002differential,ansari2017temporal}. 

The polarization degree of freedom has been the most widely deployed~\cite{liao2017satellite}, not least because polarization states are easy to generate, manipulate, and measure using high-specification off-the-shelf components. However, polarization states are not amenable to long-distance fiber transmission due to issues such as birefringence, polarization mode dispersion, polarization-dependent loss, and requirement for fast polarization compensation~\cite{liu2010decoy}; use of polarization states is, therefore, mainly restricted to free space transmission. On the other hand, time-bin states can be straightforwardly generated using fast optical modulators, transmitted through long distances in fibres and free-space~\cite{boaron2018secure,jin2019genuine}, and can support high-dimensional encoding~\cite{sasaki2014practical} leading to larger information capacity and improved noise tolerance~\cite{ecker2019overcoming}. However, their detection remains particularly challenging in terms of stability, efficiency, and flexibility. This is mainly due to the necessity of measuring time-bin superpositions, which generally requires imbalanced, or \emph{time-delayed}, interferometers~\cite{marcikic2002time,brougham2013security}. In particular, the path difference between two arms of a time-delayed interferometer is dictated by the time-bin separation time which is typically on the order of 1~ns, generally matching the timing jitter of standard single photon detectors. In practice, this amounts to a path difference of $\sim 30$~cm. Experimentally, achieving sub-wavelength interferometric phase stability over such a large path difference remains technically challenging due to various experimental disturbances, even with the use of active phase stabilization. Moreover, this active stabilization feedback loop increases the complexity of the overall system and may open the door to further attacks by an eavesdropper. Thus, the development of novel techniques with improved phase stability to measure time-bin superposition states of photons is critical to enhance the performance of time-bin quantum communication systems. 

A promising pathway towards achieving improved interferometric stability consists of reducing the path difference of the time-delayed interferometer, consequently reducing the time-bin separation time. However, this requires the use of single photon detectors with lower timing jitter. With recent developments in superconducting nanowire single photon detectors (SNSPDs), timing jitters as low as 50~ps can now be achieved in commercial devices~\cite{caloz2018high} and 3~ps in state-of-the-art devices~\cite{korzh2020demonstration}. Nevertheless, such detectors still require cooling to $\sim 1$~K, which again increases the complexity of the detection apparatus. SNSPDs have enabled the successful use of time-bins with sub-nanosecond bin width in quantum communication demonstrations~\cite{islam2017provably}. By preparing time-bins with a bin width of 400~ps, a delay interferometer can be made into a compact package where interferometric stability is achieved without active stabilization up to an hour~\cite{islam2017robust}. Nevertheless, such passive interferometric schemes still suffer from limited measurement efficiency in the superposition basis due to the non-interfering measured events.

%%%%%%%%%%%%%%%%%%%%%%%%%
\begin{figure}[t!]
	\centering
		\includegraphics[width=0.44\textwidth]{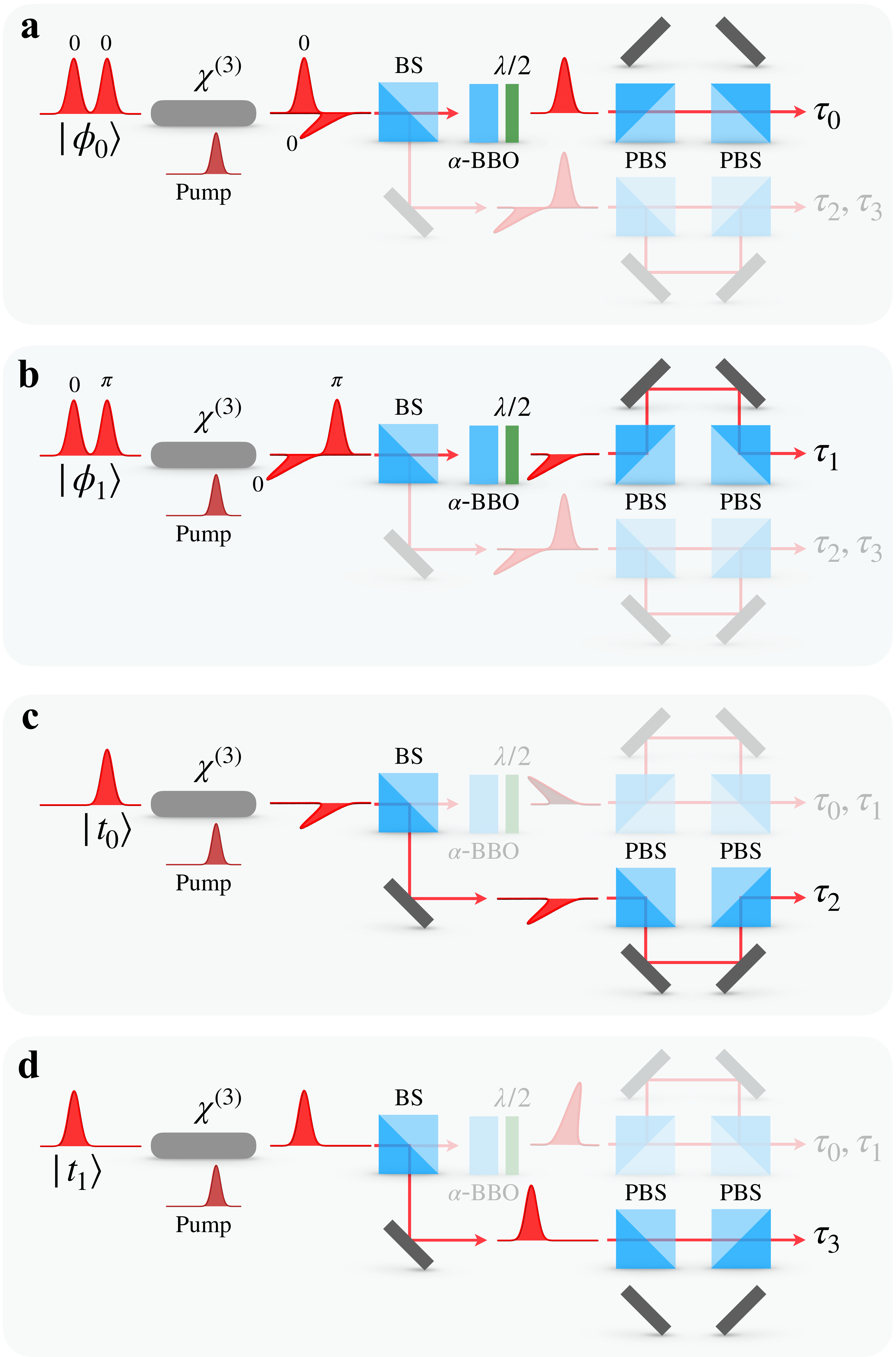}
	\caption{\textbf{Time-bin states measurement}. Conceptual experimental setup for \textbf{(a,b)} the phase basis measurement and \textbf{(c,d)} the time basis measurement. A strong pump is overlapped with the $t_0$ time-bin inside a $\chi^{(3)}$ material rotating its polarization. A 50:50 beam splitter randomly selects one of two measurement settings, i.e. the phase basis (upper) or time basis (lower). In the scenario where the input state is prepared in one basis but measured in the other basis (shaded part of the setup), the measurement outcomes are random and will be later sifted. $\chi^{(3)}$, third-order nonlinear material; BS, beam splitter; $\lambda/2$, half-wave plate; PBS, polarizing beam splitter.    
	}
	\label{fig:concept}
\end{figure}
%%%%%%%%%%%%%%%%%%%%%%%%%

%%%%%%%%%%%%%%%%%%%%%%%%%
\begin{figure*}[t!]
	\centering
		\includegraphics[width=0.75\textwidth]{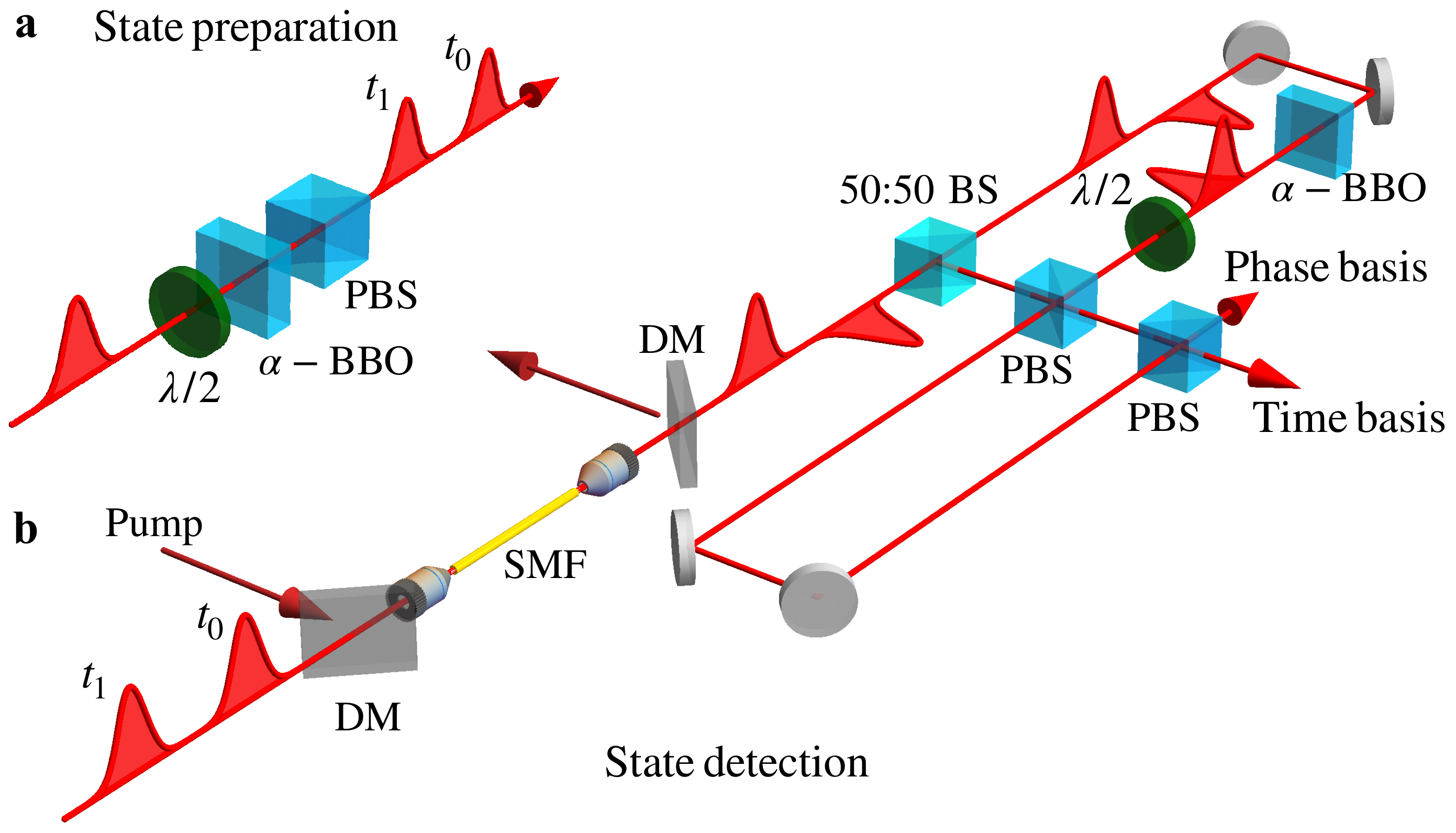}
	\caption{\textbf{Experimental setup}. Simplified experimental setup showing the state preparation and the state detection of ultrafast time-bin qubit states in a BB84 protocol experiment. \textbf{(a)} In the state preparation stage, an ultrashort pulse is sent to a 10-mm long $\alpha$-BBO birefringent crystal splitting the input pulse into two output pulses with orthogonal polarization separated by 4.5~ps. A polarizing beam splitter (PBS) is then used to erase the polarization information associated with each time-bin resulting in a uniformly polarized time-bin state. The weight and relative phase of the time-bin state can be determined by a half-wave plate ($\lambda/2$) placed before the $\alpha$-BBO. \textbf{(b)} In the detection stage, the signal pulse is combined with a strong pump pulse at a dichroic mirror (DM) and then coupled to a single mode fiber where the pump pulse switches the early time-bin. The pump pulse is then filtered out using a DM and further spectral filters not shown in the figure. A 50:50 beam splitter (BS) is used to randomly switch the measurement setting from the time basis (reflected output) to the phase basis (transmitted output). In the time basis a delayed-interferometer is designed with a path difference of 88~cm such that the time-bin state can be detected by measuring the time-of-arrival of the pulse with standard single photon detectors. In the phase basis, a second 10-mm $\alpha$-BBO crystal is used to recombine the pulses and converts the relative phase information to polarization. The same delayed-interferometer can be used in parallel where the phase state is determined from the time-of-arrival of detected pulses. }
	\label{fig:setup}
\end{figure*}
%%%%%%%%%%%%%%%%%%%%%%%%%

Recent developments in ultrafast quantum optics~\cite{maclean2018direct,england2015storage} adds to the current experimental toolbox for photonic time-based quantum information processing. Shifting time-bin qubits to the realm of ultrafast pulses allows the bin-widths to be compressed as low as a few picoseconds. Time-delayed interferometers then only require path differences of a few hundred micrometers, offering the potential for intrinsic passive interferometric stability over long periods of time. Here, we propose and experimentally perform a proof-of-principle quantum communication experiment using ultrafast time bins~\cite{kupchak2017time}, with a bin-width of $\Delta \tau=4.5$~ps and complete encoding of the qubit in a 7~ps window. The time-delayed interferometers required first for qubit creation, and later for measurement in the phase basis, are each based on birefringent crystals. The two arms of each interferometer are collinear and their path difference is due to the difference in group index experienced by pulses with orthogonal polarizations in the crystal. For example, a 10-mm-long $\alpha$-BBO crystal can induce a time delay of $\Delta\tau=4.5$~ps between orthogonally polarized pulses at 720.8~nm. Such common-path interferometers have excellent passive phase stability, and require no active phase compensation. The ultrafast bin-widths are significantly smaller than the timing jitter of conventional single photon detectors. The time-of-arrival of the time-bin states is therefore measured by ultrafast polarization switching using cross-phase modulation in a single mode fiber, applied with an intense ultrafast control pulse~\cite{kupchak2017time,kupchak2019terahertz}, see Fig.~\ref{fig:concept}. Moreover, in the superposition (phase) basis, we take advantage of the cross-phase modulation polarization switching to deliver a measurement efficiency for our scheme of 100~\%, in theory, compared to a measurement efficiency of 50~\% for standard time-delayed interferometery measurements where non-interfering events gives no information about the relative phase of time-bin superpositions.

\section{Experiment}

We demonstrate the experimental feasibility of ultrafast time-bin qubits for quantum communication by performing a proof-of-principle experiment of a time-bin-based decoy-state BB84 protocol. Our experiment consists of a sender, \emph{Alice}, and a receiver, \emph{Bob}. Alice prepares weak coherent pulses (WCPs) by attenuating ultrafast pulses to the single-photon level. The mean photon number of the WCPs is optimized given channel conditions, e.g., channel loss and quantum bit error rate (QBER). The pulses are obtained from an optical parametric oscillator pumped by a Ti:sapphire laser at a repetition rate of $f_\mathrm{rep}=80$~MHz. Pulses are generated at a central wavelength of $\lambda_\mathrm{signal}=720.8$~nm with a spectral bandwidth of $\Delta \lambda_\mathrm{signal}=1.7$~nm full-width at half maximum. In the original BB84 protocol, Alice randomly selects one of four polarization states and transmits the encoded photons to Bob. These four states belong to two mutually unbiased bases (MUBs), i.e., a computational basis and a superposition basis. For time-bins, these two bases are usually referred to as the \emph{time basis} and the \emph{phase basis}, where the former contains the states $|t_0\rangle$ and $|t_1\rangle$, and the latter contains the states $|\phi_0\rangle=1/\sqrt{2} \left( |t_0\rangle + |t_1\rangle \right)$ and $|\phi_1\rangle=1/\sqrt{2} \left( |t_0\rangle - |t_1\rangle \right)$, respectively. Using a half-wave plate (HWP), an $\alpha$-BBO crystal and a polarizing beam splitter (PBS), Alice generates all four time-bin states by varying the angle of the HWP to $0^\circ$, $45^\circ$, $-22.5^\circ$, and $22.5^\circ$, see Fig.~\ref{fig:setup}-\textbf{a}.

%%%%%%%%%%%%%%%%%%%%%%%%%
\begin{figure}[t!]
	\centering
		\includegraphics[width=0.44\textwidth]{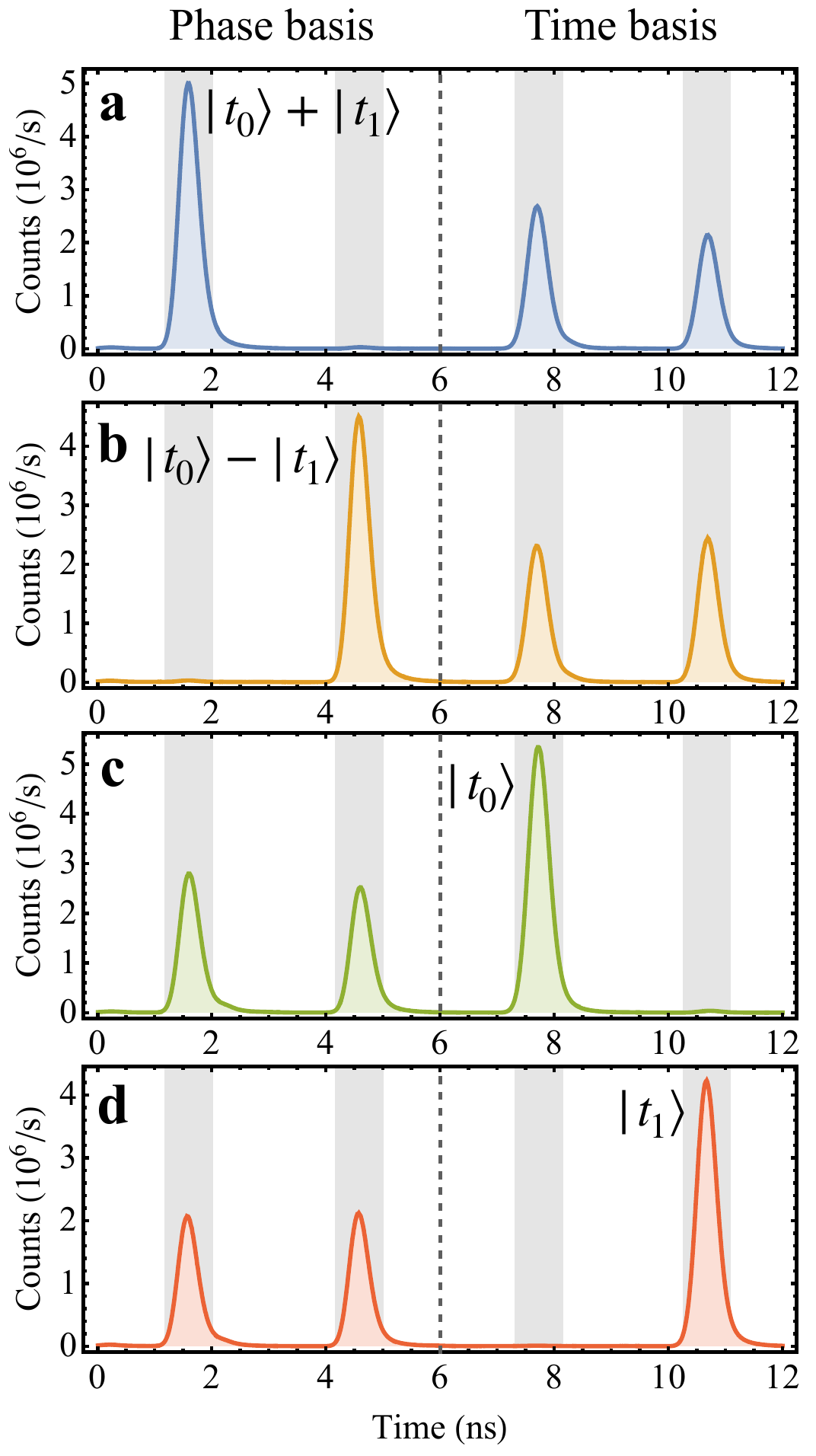}
	\caption{\textbf{Time-bin states measurement}. \textbf{(a,b)} Phase basis measurement. The single photon data. \textbf{(c,d)} Time basis measurement. The shaded area represents to the temporal window corresponding to each time-bins state. The number of counts shown are taken for an integration time of 1 second. The gray dotted line show the separation between the phase basis and the time basis. 
	}
	\label{fig:timebin}
\end{figure}
%%%%%%%%%%%%%%%%%%%%%%%%%

Upon receiving the transmitted photons, Bob measures the time-bin states and generates a raw key by assigning a value of 0 to the measured state if it is found to be in the state $|t_0\rangle$ or $|\phi_0\rangle$, and a value of 1 when found to be in $|t_1\rangle$ or $|\phi_1\rangle$. In order to measure the ultrafast time-bin states, a synchronized pump pulse is combined with the signal pulses using a dichroic mirror (DM). The signal and pump pulses are then coupled to a single mode fiber (SMF), with a coupling efficiency of 50~\% and 65~\%, respectively. The pump pulses are prepared at a center wavelength of $\lambda_\mathrm{pump}=800$~nm and are spectrally filtered with a pair of angle-tuned bandpass filters such that $\Delta \lambda_\mathrm{pump}=2.1$~nm. In the presence of the strong pump pulse, the third-order nonlinearity of the SMF is used to induce a birefringence in the SMF causing a rotation of the polarization of the signal pulses at time $t_0$ using cross-phase modulation via the optical Kerr effect. We note that for time-bin superposition states, the relative phase between the states $|t_0\rangle$ and $|t_1\rangle$ is preserved up to an additional constant phase introduced by the pump during the polarization rotation. It has been shown that such an optical switch can rotate the polarization of single-photon pulses with unit efficiency while introducing very little noise~\cite{kupchak2019terahertz}. The switching efficiency, $\eta$, of the signal pulses is given by 
\begin{eqnarray}
\eta=\sin^2 \left(2 \theta \right) \sin^2 \left(\frac{\Delta \phi}{2} \right),
\end{eqnarray}
where $\theta$ is the angle between the polarization of the signal and pump pulses, ${\Delta \phi=8 \pi n_2 L_\mathrm{eff}/3 \lambda_\mathrm{signal}}$ is the induced nonlinear phase shift in SMF in the presence of the pump pulse, $n_2$ is the nonlinear refractive index of the SMF, $L_\mathrm{eff}$ is the effective length of the nonlinear medium, and $I_\mathrm{pump}$ is the intensity of the pump. A unit switching efficiency is achieved by setting $\theta=\pi/4$, $\Delta \phi = \pi$, and by taking advantage of the group velocity mismatch between the signal and pump wavelength in the SMF to achieve a full temporal walkoff of the signal and the pump pulses.

%%%%%%%%%%%%%%%%%%%%%%%%%
\begin{figure}[t!]
	\centering
		\includegraphics[width=0.49\textwidth]{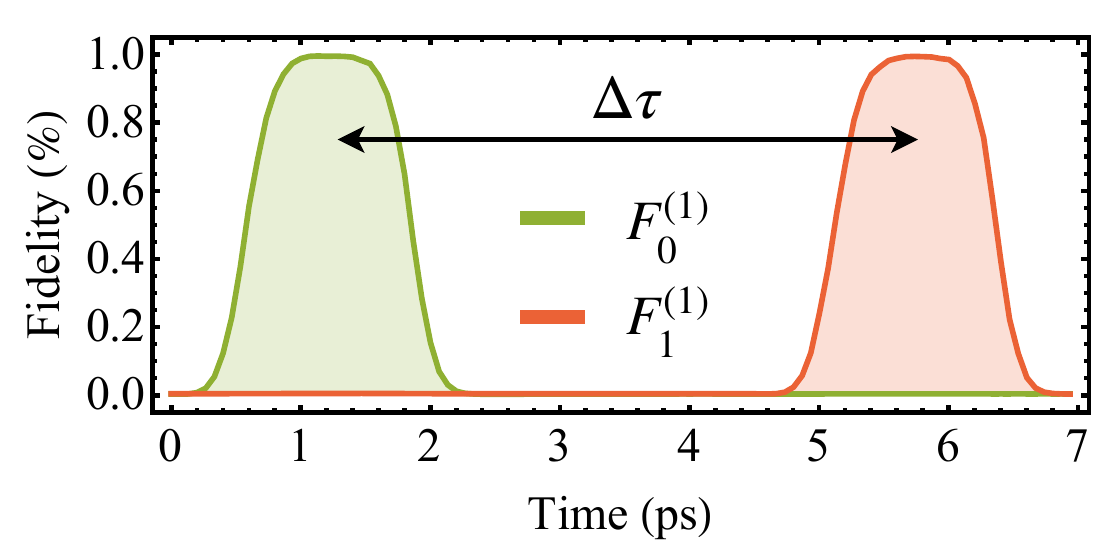}
	\caption{\textbf{Pump delay measurements.}. Measurement of the state fidelity for the time basis, i.e. $F^{(1)}_0$ (solid green curve) and $F^{(1)}_1$ (solid red curve). The arrow shows the time-bin separation, or bin-width, $\Delta \tau$. The shape of the fidelity curve as a function of pump delay is dictated by the pulse duration of the pump and signal, and by the difference in group velocity of the pump and signal inside the SMF. 
	}
	\label{fig:pump}
\end{figure}
%%%%%%%%%%%%%%%%%%%%%%%%%

After mapping the ultrafast time bins on to orthogonal polarization states, we use a polarizing delayed-interferometer with a path difference of 88 cm to map the distinct polarization states to separate nanosecond time-bins, which can then be straightforwardly detected by avalanche photodiodes (APDs) and processed using a time-to-digital converter (TDC) (Swabian Instruments, Time Tagger Ultra). As is required in the BB84 protocol, Bob randomly varies the measurement basis between the time basis and the phase basis: a 50:50 beam splitter (BS) directs the photons randomly to the two alternative measurement pathways (see Fig. 2-b) through the shared polarizing delayed-interferometer. In the time-basis pathway, propagation through the polarizing delayed-interferometer maps the two orthogonal polarizations directly to distinct time bins at the time-basis exit port. In the phase-basis pathway, the time difference between the orthogonal polarization states is initially compensated using an $\alpha$-BBO crystal, with the resulting interference mapping the phase information to polarization. This compact setup delivers a high phase stability for phase basis measurements. The polarizing delayed-interferometer then maps the two orthogonal polarizations directly to distinct time bins at the phase-basis exit port. Sharing the polarizing delayed intereferometer for the time and phase bases allows measurement of the four distinct states with only two APDs, while maintaining a compact footprint. We note that the polarizing delayed-interferometer is only used to measure polarization and does not require phase stability since no interference occurs at the output ports. The final nanosecond signal pulses are analyzed to determine the state sent from Alice, see Fig.~\ref{fig:timebin}. In particular, a temporal window of 0.8~ns is defined for each time-bin state. The probability of detection $P_{i,j}^{(\alpha,\beta)}=\left|\langle \psi^{(\beta)}_j | \psi^{(\alpha)}_i \rangle \right|^2$ is experimentally determined, where $|\psi^{(0)}_0\rangle=|\phi_0\rangle$, $|\psi^{(0)}_1\rangle=|\phi_1\rangle$, $|\psi^{(1)}_0\rangle=|t_0\rangle$, and $|\psi^{(1)}_1\rangle=|t_1\rangle$. The measurement bases $\alpha$ and $\beta$ correspond to Alice's preparation basis and Bob's measurement basis, respectively. The probability of detection is obtained from measured detection events, i.e. $P_{i,j}^{(\alpha,\beta)}=N^{(\alpha,\beta)}_{i,j}/\sum_{k=0}^1 N^{(\alpha,\beta)}_{i,k}$. The state fidelity is then obtained from the probability of detection, i.e. $F^{(\alpha)}_i=P^{(\alpha,\alpha)}_{i,i}$.

We further characterize our experimental setup by varying the delay between the pump and signal pulses while measuring each output state. By repeating these measurements for all input states, we can extract information about the pump pulse temporal profile and the temporal walkoff of the signal and pump pulses. The state fidelity in the time basis as a function of pump delay is shown in Fig.~\ref{fig:pump}. Moreover, we measure the temporal separation between the time states $|t_0\rangle$ and $|t_1\rangle$ to be $\Delta \tau=4.5$~ps, in agreement with the expected value from the $\alpha$-BBO crystal's thickness.  

%%%%%%%%%%%%%%%%%%%%%%%%%
\begin{figure*}[ht]
	\centering
		\includegraphics[width=0.78\textwidth]{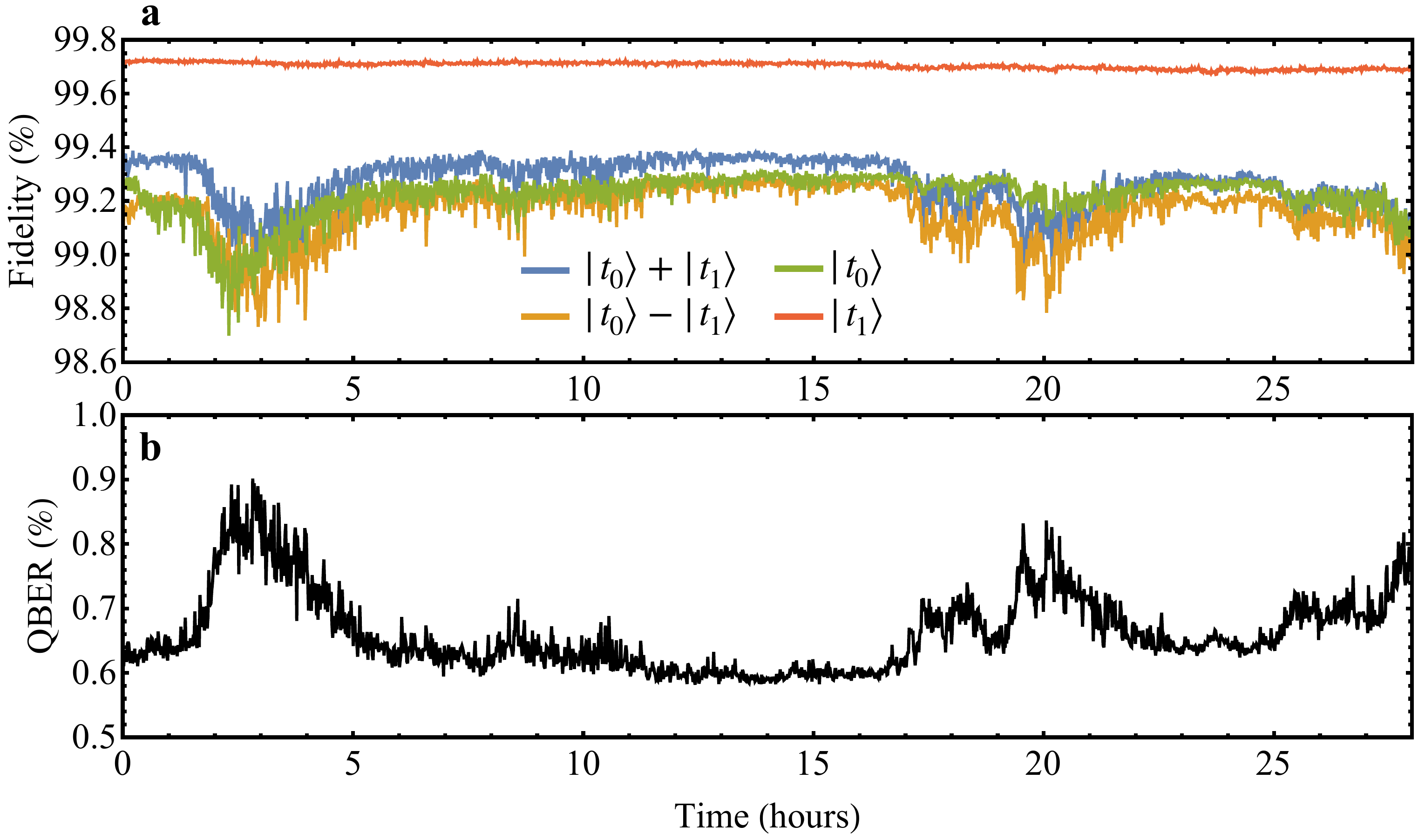}
	\caption{\textbf{Time-bin measurement over 28 hours}. \textbf{(a)}. The state fidelity for all four measured states, i.e., $|\phi_0\rangle$ (blue curve), $|\phi_1\rangle$ (yellow curve), $|t_0\rangle$ (green curve), $|t_1\rangle$ (red curve). \textbf{(b)} The QBER is calculated from the state fidelity averaged over all states. 
	}
	\label{fig:hours}
\end{figure*}
%%%%%%%%%%%%%%%%%%%%%%%%%

%%%%%%%%%%%%%%%%%%%%%%%%%
\begin{figure}[ht]
	\centering
		\includegraphics[width=0.35\textwidth]{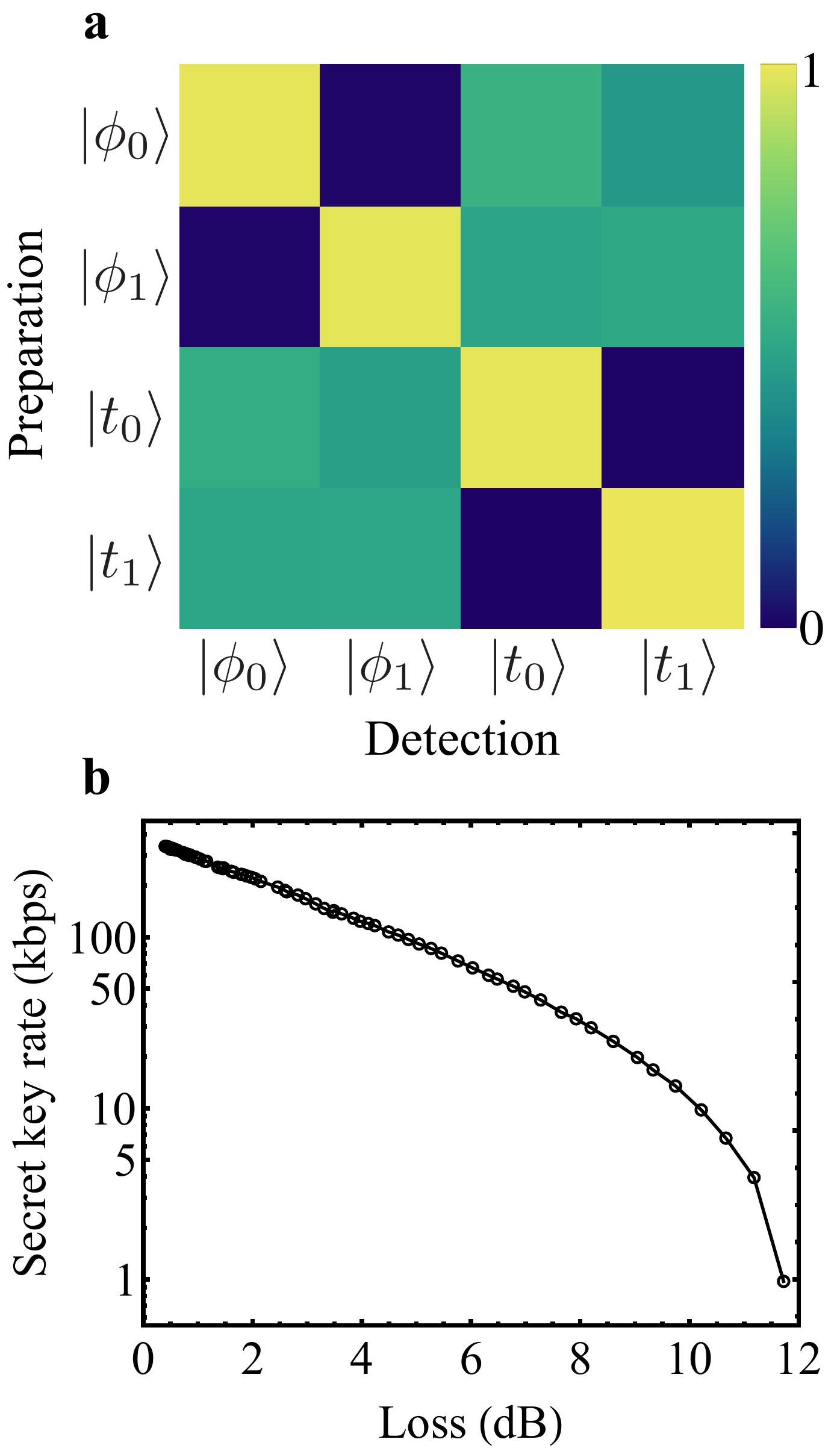}
	\caption{\textbf{QKD results.} \textbf{(a)} Probability-of-detection matrix. The rows correspond to different prepared input time-bin states and the columns correspond to different measured output states. \textbf{(b)} Secret key rate as a function of channel loss.
	}
	\label{fig:qkd}
\end{figure}
%%%%%%%%%%%%%%%%%%%%%%%%%

\section{Results}

We now test the experimental viability of ultrafast time-bin qubits in the context of quantum communication. In particular, we use our generation and detection experimental setting to perform a proof-of-principle QKD demonstration using the formalism of the decoy-state BB84 protocol where the secret key rate is the key metric of performance. By doing so, we will experimentally demonstrate that ultrafast time-bins offer the unique advantage of extended intrinsic phase stability leading to a QBER lower than 1~\% for time durations as high as 28 hours, while delivering a high detection efficiency in both measurement bases, which are key requirements in quantum communication. The performance of our experimental setup is evaluated quantitavely using the standard decoy BB84 post-processing procedure~\cite{ma2005practical}, where the following formula is used to determine key generation rates:
\begin{eqnarray}
R \geq q \left( -Q_\mu f\left(E_\mu \right) H_2\left( E_\mu \right) + Q_1 \left[ 1-H_2\left( e_1 \right) \right] \right),
\end{eqnarray}
where $q=1/2$ is the sifting efficiency, $Q_\mu$ is the gain of signal states, $f(x)$ is the error correction efficiency, ${H_2(x)=-x \log_2(x)-(1-x) \log_2(1-x)}$ is the binary Shannon entropy function, $E_\mu$ is the QBER, $Q_1$ is the gain of single-photon states, and $e_1$ is the error rate of single-photon states. The QBER can be directly obtained from state fidelities, i.e. ${E_\mu=1-\frac{1}{4}\sum_{\alpha=0,i=0}^{1,1} F^{(\alpha)}_i}$. A standard error correction efficiency factor of $f\left(E_\mu \right)=1.22$ is used in our analysis.

As a first test, we measure the state fidelity for each input time-bin state over the course of several hours, see Fig.~\ref{fig:hours}-\textbf{a}. By doing so we demonstrate the feasibility of using time-bin states in quantum communication where time-delayed interferometry is inherently phase-stable due to the small path difference and common-path nature of our interferometer. We show an average state fidelity in excess of 99~\%, corresponding to a QBER below 1~\%, see Fig.~\ref{fig:hours}. Variations in state fidelities over time can be attributed to slight fluctuations in the intensity or polarization of the pump pulses. We note that the time-bin state $|t_1\rangle$ has a much less variable state fidelity since it is unaffected by the pump pulse. From the measurements in Fig.~\ref{fig:hours}-\textbf{a}, we can calculate a probability-of-detection matrix which is taken over the integration time of 28~hours, see Fig.~\ref{fig:qkd}-\textbf{a}. Measurements done in different MUBs yield random results which are subsequently discarded in the sifting phase of the BB84 protocol. From the averaged probability-of-detection matrix, we obtain an averaged QBER of ${E_\mu=(0.8\pm0.3)~\%}$ over the integration time of 28~hours. The secret key rate, $R$, can then be calculated over this period of time. 

The mean photon number of the signal pulse is set to $\mu=0.8$ and the brightness of the decoy states $\nu$ is optimized depending on channel conditions. This is achieved using a combination of neutral density filters, a HWP mounted in a motorized stage, and a PBS. The modulation of the of the signal and decoy states guarantees the security of our demonstration against photon number splitting attacks. An additional pair of HWP and PBS is used to introduce losses in the channel. The secret key rate is then measured for channel loss conditions varying from 0.45 to 12~dB channel loss. Hence, the total loss of our system is varied from 14.6 to 26.1~dB, with an additional 3~dB SMF coupling loss, 2.2~dB detector efficiency loss, and another 8.9~dB loss at Bob's detection stage (waveplates, PBS, $\alpha$-BBO, and spectral filters). For a total loss of 14.6~dB, we measure a secret key rate of 0.34~Mbps.

\section{Discussion and Outlook}

 From our QKD demonstration, we have experimentally demonstrated that ultrafast time-bin qubits offer a versatile platform for QKD where standard protocols such as the BB84 protocol can now be considered due to the ability to directly measure the time-bin states in MUBs with high detection efficiencies. Previously, the temporal degree of freedom of photons has been an important candidate in quantum communication, particularly due to its ease of use in fiber networks. However, due to the lack of efficient, stable, and passive measurement techniques for time-bin superposition states, more complicated QKD protocols are yet to be employed. For example, the differential phase shift~\cite{inoue2002differential}, the coherent one way~\cite{stucki2005fast}, and the round-robin differential phase shift~\cite{sasaki2014practical} protocols were specifically design to overcome the technical difficulties associated with measuring time-bin qubits. On the other hand, the BB84 protocol, which was initially designed for polarization states, offers a higher overall efficiency, a simpler post-processing, and has well-studied security considerations in practical implementations~\cite{xu2020secure}.

Another QKD protocol with a high efficiency is the six-states protocol~\cite{bruss1998optimal}, where an additional MUB is considered. Though reducing the sifting efficiency, the tomographic nature of the protocol results in a slightly larger error threshold which can translate to a larger amount of loss tolerability. We note that our experimental setup is also suited to perform the six states protocol, where a quarter-wave plate is added to both the preparation and detection stages. This will result in the generation and measurement of a third MUB consisting of the states $(|t_0\rangle \pm i |t_1\rangle)/\sqrt{2}$. The addition of the third MUB would allow the tomographic reconstruction of ultrafast time-bin qubits.

In using ultrafast pulses, we are taking advantage of the favourable properties of near-Fourier-transform-limited pulses to encode information onto photons in the most condensed manner possible. Our expriment uses signal pulses that are in a near-single spectro-temporal mode. In the ultrafast regime, we have a direct access to both the temporal and spectral domains. This feature, that is notable with ultrafast pulses, can be exploited to achieve passive noise filtering on both the temporal and spectral degrees of freedom. By doing so, a noise tolerance approaching the ultimate limit in QKD can be achieved~\cite{bouchard2021achieving,raymer2020time}, this approach is compatible with the use of ultrafast time-bin qubits. Moreover, mature technologies exist to compensate for channel disturbances such as chromatic dispersion occurring in long-distance fiber networks. Finally, we note that our experimental platform can be extended to more general time-bin states such as two-photon entangled states and high-dimensional time-bin states, also known as \emph{qudits}. Our technique using cross-phase modulation inside an SMF via the optical Kerr effect is compatible with the measurement of entangled and qudit states leading to the development of a larger toolkit to perform quantum communication in the time domain.\\

In conclusion, we experimentally demonstrated the use of ultrafast time-bin qubits in quantum communication by experimentally investigating the performance of a proof-of-principle decoy-state BB84 QKD protocol. By significantly reducing the size of the time bins from nanoseconds to picoseconds, we enable the use of time-delayed interferometers with excellent inherent phase stability. Here, this is achieved using a 10-mm thick birefringent crystal as a common-path time-delayed interferometer. Thus, an average state fidelity in excess of 99~\% is achieved over a period of time of 28~hours, resulting in a secret key rate of 0.34~Mbps over this time duration in low-loss conditions. These results provides a new platform that is compatible with quantum communication systems through fiber networks or free space. \\

\section*{Acknowledgements}
This work is supported by the High Throughput Secure Networks Challenge Program at the National Research Council of Canada, the Natural Sciences and Engineering Research Council of Canada, and the University of Ottawa-NRC Joint Centre for Extreme Photonics. The authors thank Ebrahim Karimi, Rune Lausten, Denis Guay, Doug Moffatt, Kent Bonsma-Fisher, and Kate Fenwick for support and insightful discussions.

%\bibliography{ultrafastQKD}

\begin{thebibliography}{29}%
\makeatletter
\providecommand \@ifxundefined [1]{%
 \@ifx{#1\undefined}
}%
\providecommand \@ifnum [1]{%
 \ifnum #1\expandafter \@firstoftwo
 \else \expandafter \@secondoftwo
 \fi
}%
\providecommand \@ifx [1]{%
 \ifx #1\expandafter \@firstoftwo
 \else \expandafter \@secondoftwo
 \fi
}%
\providecommand \natexlab [1]{#1}%
\providecommand \enquote  [1]{``#1''}%
\providecommand \bibnamefont  [1]{#1}%
\providecommand \bibfnamefont [1]{#1}%
\providecommand \citenamefont [1]{#1}%
\providecommand \href@noop [0]{\@secondoftwo}%
\providecommand \href [0]{\begingroup \@sanitize@url \@href}%
\providecommand \@href[1]{\@@startlink{#1}\@@href}%
\providecommand \@@href[1]{\endgroup#1\@@endlink}%
\providecommand \@sanitize@url [0]{\catcode `\\12\catcode `\$12\catcode
  `\&12\catcode `\#12\catcode `\^12\catcode `\_12\catcode `\%12\relax}%
\providecommand \@@startlink[1]{}%
\providecommand \@@endlink[0]{}%
\providecommand \url  [0]{\begingroup\@sanitize@url \@url }%
\providecommand \@url [1]{\endgroup\@href {#1}{\urlprefix }}%
\providecommand \urlprefix  [0]{URL }%
\providecommand \Eprint [0]{\href }%
\providecommand \doibase [0]{http://dx.doi.org/}%
\providecommand \selectlanguage [0]{\@gobble}%
\providecommand \bibinfo  [0]{\@secondoftwo}%
\providecommand \bibfield  [0]{\@secondoftwo}%
\providecommand \translation [1]{[#1]}%
\providecommand \BibitemOpen [0]{}%
\providecommand \bibitemStop [0]{}%
\providecommand \bibitemNoStop [0]{.\EOS\space}%
\providecommand \EOS [0]{\spacefactor3000\relax}%
\providecommand \BibitemShut  [1]{\csname bibitem#1\endcsname}%
\let\auto@bib@innerbib\@empty
%</preamble>
\bibitem [{\citenamefont {Scarani}\ \emph {et~al.}(2009)\citenamefont
  {Scarani}, \citenamefont {Bechmann-Pasquinucci}, \citenamefont {Cerf},
  \citenamefont {Du\ifmmode~\check{s}\else \v{s}\fi{}ek}, \citenamefont
  {L\"utkenhaus},\ and\ \citenamefont {Peev}}]{scarani:09}%
  \BibitemOpen
  \bibfield  {author} {\bibinfo {author} {\bibfnamefont {V.}~\bibnamefont
  {Scarani}}, \bibinfo {author} {\bibfnamefont {H.}~\bibnamefont
  {Bechmann-Pasquinucci}}, \bibinfo {author} {\bibfnamefont {N.~J.}\
  \bibnamefont {Cerf}}, \bibinfo {author} {\bibfnamefont {M.}~\bibnamefont
  {Du\ifmmode~\check{s}\else \v{s}\fi{}ek}}, \bibinfo {author} {\bibfnamefont
  {N.}~\bibnamefont {L\"utkenhaus}}, and\ \bibinfo {author} {\bibfnamefont
  {M.}~\bibnamefont {Peev}},\ }\newblock \bibinfo{title}{The security of practical quantum key distribution,} \href {https://journals.aps.org/rmp/abstract/10.1103/RevModPhys.81.1301}
  {\bibfield  {journal} {\bibinfo  {journal} {Rev. Mod. Phys.}\ }\textbf
  {\bibinfo {volume} {81}},\ \bibinfo {pages} {1301} (\bibinfo {year}
  {2009})}\BibitemShut {NoStop}%
\bibitem [{\citenamefont {Xu}\ \emph {et~al.}(2020)\citenamefont {Xu},
  \citenamefont {Ma}, \citenamefont {Zhang}, \citenamefont {Lo},\ and\
  \citenamefont {Pan}}]{xu2020secure}%
  \BibitemOpen
  \bibfield  {author} {\bibinfo {author} {\bibfnamefont {F.}~\bibnamefont
  {Xu}}, \bibinfo {author} {\bibfnamefont {X.}~\bibnamefont {Ma}}, \bibinfo
  {author} {\bibfnamefont {Q.}~\bibnamefont {Zhang}}, \bibinfo {author}
  {\bibfnamefont {H.-K.}\ \bibnamefont {Lo}}, and\ \bibinfo {author}
  {\bibfnamefont {J.-W.}\ \bibnamefont {Pan}},} \newblock \bibinfo{title}{Secure quantum key distribution with realistic devices,} \href{https://journals.aps.org/rmp/abstract/10.1103/RevModPhys.92.025002}{\bibfield
  {journal} {\bibinfo  {journal} {Rev. Mod. Phys.}\ }\textbf
  {\bibinfo {volume} {92}},\ \bibinfo {pages} {025002} (\bibinfo {year}
  {2020})}\BibitemShut {NoStop}%
\bibitem [{\citenamefont {Ding}\ \emph {et~al.}(2017)\citenamefont {Ding},
  \citenamefont {Bacco}, \citenamefont {Dalgaard}, \citenamefont {Cai},
  \citenamefont {Zhou}, \citenamefont {Rottwitt},\ and\ \citenamefont
  {Oxenl{\o}we}}]{ding2017high}%
  \BibitemOpen
  \bibfield  {author} {\bibinfo {author} {\bibfnamefont {Y.}~\bibnamefont
  {Ding}}, \bibinfo {author} {\bibfnamefont {D.}~\bibnamefont {Bacco}},
  \bibinfo {author} {\bibfnamefont {K.}~\bibnamefont {Dalgaard}}, \bibinfo
  {author} {\bibfnamefont {X.}~\bibnamefont {Cai}}, \bibinfo {author}
  {\bibfnamefont {X.}~\bibnamefont {Zhou}}, \bibinfo {author} {\bibfnamefont
  {K.}~\bibnamefont {Rottwitt}}, \ and\ \bibinfo {author} {\bibfnamefont
  {L.~K.}\ \bibnamefont {Oxenl{\o}we}},\ } \newblock \bibinfo{title}{High-dimensional quantum key distribution based on multicore fiber using silicon photonic integrated circuits,} \href{https://www.nature.com/articles/s41534-017-0026-2}{\bibfield  {journal}
  {\bibinfo  {journal} {Npj Quantum Inf.}\ }\textbf {\bibinfo {volume}
  {3}},\ \bibinfo {pages} {25} (\bibinfo {year} {2017})}\BibitemShut {NoStop}%
\bibitem [{\citenamefont {Mirhosseini}\ \emph {et~al.}(2015)\citenamefont
  {Mirhosseini}, \citenamefont {Maga{\~n}a-Loaiza}, \citenamefont {O'Sullivan},
  \citenamefont {Rodenburg}, \citenamefont {Malik}, \citenamefont {Lavery},
  \citenamefont {Padgett}, \citenamefont {Gauthier},\ and\ \citenamefont
  {Boyd}}]{mirhosseini:15}%
  \BibitemOpen
  \bibfield  {author} {\bibinfo {author} {\bibfnamefont {M.}~\bibnamefont
  {Mirhosseini}}, \bibinfo {author} {\bibfnamefont {O.~S.}\ \bibnamefont
  {Maga{\~n}a-Loaiza}}, \bibinfo {author} {\bibfnamefont {M.~N.}\ \bibnamefont
  {O'Sullivan}}, \bibinfo {author} {\bibfnamefont {B.}~\bibnamefont
  {Rodenburg}}, \bibinfo {author} {\bibfnamefont {M.}~\bibnamefont {Malik}},
  \bibinfo {author} {\bibfnamefont {M.~P.}\ \bibnamefont {Lavery}}, \bibinfo
  {author} {\bibfnamefont {M.~J.}\ \bibnamefont {Padgett}}, \bibinfo {author}
  {\bibfnamefont {D.~J.}\ \bibnamefont {Gauthier}}, and\ \bibinfo {author}
  {\bibfnamefont {R.~W.}\ \bibnamefont {Boyd}},\ } \newblock \bibinfo{title}{High-dimensional quantum cryptography with twisted light,} \href{https://iopscience.iop.org/article/10.1088/1367-2630/17/3/033033} {\bibfield
  {journal} {\bibinfo  {journal} {New J. Phys.}\ }\textbf {\bibinfo
  {volume} {17}},\ \bibinfo {pages} {033033} (\bibinfo {year}
  {2015})}\BibitemShut {NoStop}%
\bibitem [{\citenamefont {Bouchard}\ \emph {et~al.}(2018)\citenamefont
  {Bouchard}, \citenamefont {Heshami}, \citenamefont {England}, \citenamefont
  {Fickler}, \citenamefont {Boyd}, \citenamefont {Englert}, \citenamefont
  {S{\'{a}}nchez-Soto},\ and\ \citenamefont
  {Karimi}}]{bouchard2018experimental}%
  \BibitemOpen
  \bibfield  {author} {\bibinfo {author} {\bibfnamefont {F.}~\bibnamefont
  {Bouchard}}, \bibinfo {author} {\bibfnamefont {K.}~\bibnamefont {Heshami}},
  \bibinfo {author} {\bibfnamefont {D.}~\bibnamefont {England}}, \bibinfo
  {author} {\bibfnamefont {R.}~\bibnamefont {Fickler}}, \bibinfo {author}
  {\bibfnamefont {R.~W.}\ \bibnamefont {Boyd}}, \bibinfo {author}
  {\bibfnamefont {B.-G.}\ \bibnamefont {Englert}}, \bibinfo {author}
  {\bibfnamefont {L.~L.}\ \bibnamefont {S{\'{a}}nchez-Soto}}, and\ \bibinfo
  {author} {\bibfnamefont {E.}~\bibnamefont {Karimi}},\ } \newblock \bibinfo{title}{Experimental investigation of high-dimensional quantum key distribution protocols with twisted photons,} \href{\doibase
  10.22331/q-2018-12-04-111} {\bibfield  {journal} {\bibinfo  {journal}
  {{Quantum}}\ }\textbf {\bibinfo {volume} {2}},\ \bibinfo {pages} {111}
  (\bibinfo {year} {2018})}\BibitemShut {NoStop}%
\bibitem [{\citenamefont {Mower}\ \emph {et~al.}(2013)\citenamefont {Mower},
  \citenamefont {Zhang}, \citenamefont {Desjardins}, \citenamefont {Lee},
  \citenamefont {Shapiro},\ and\ \citenamefont {Englund}}]{mower2013high}%
  \BibitemOpen
  \bibfield  {author} {\bibinfo {author} {\bibfnamefont {J.}~\bibnamefont
  {Mower}}, \bibinfo {author} {\bibfnamefont {Z.}~\bibnamefont {Zhang}},
  \bibinfo {author} {\bibfnamefont {P.}~\bibnamefont {Desjardins}}, \bibinfo
  {author} {\bibfnamefont {C.}~\bibnamefont {Lee}}, \bibinfo {author}
  {\bibfnamefont {J.~H.}\ \bibnamefont {Shapiro}}, \ and\ \bibinfo {author}
  {\bibfnamefont {D.}~\bibnamefont {Englund}},\ } \newblock \bibinfo{title}{High-dimensional quantum key distribution using dispersive optics,} \href{https://journals.aps.org/pra/abstract/10.1103/PhysRevA.87.062322}{\bibfield
  {journal} {\bibinfo  {journal} {Phys. Rev. A}\ }\textbf {\bibinfo
  {volume} {87}},\ \bibinfo {pages} {062322} (\bibinfo {year}
  {2013})}\BibitemShut {NoStop}%
\bibitem [{\citenamefont {Inoue}\ \emph {et~al.}(2002)\citenamefont {Inoue},
  \citenamefont {Waks},\ and\ \citenamefont
  {Yamamoto}}]{inoue2002differential}%
  \BibitemOpen
  \bibfield  {author} {\bibinfo {author} {\bibfnamefont {K.}~\bibnamefont
  {Inoue}}, \bibinfo {author} {\bibfnamefont {E.}~\bibnamefont {Waks}}, \ and\
  \bibinfo {author} {\bibfnamefont {Y.}~\bibnamefont {Yamamoto}},\ } \newblock \bibinfo{title}{Differential Phase Shift Quantum Key Distribution,} \href{https://journals.aps.org/prl/abstract/10.1103/PhysRevLett.89.037902} {\bibfield  {journal} {\bibinfo  {journal} {Phys. Rev. Lett.}\
  }\textbf {\bibinfo {volume} {89}},\ \bibinfo {pages} {037902} (\bibinfo
  {year} {2002})}\BibitemShut {NoStop}%
\bibitem [{\citenamefont {Ansari}\ \emph {et~al.}(2017)\citenamefont {Ansari},
  \citenamefont {Harder}, \citenamefont {Allgaier}, \citenamefont {Brecht},\
  and\ \citenamefont {Silberhorn}}]{ansari2017temporal}%
  \BibitemOpen
  \bibfield  {author} {\bibinfo {author} {\bibfnamefont {V.}~\bibnamefont
  {Ansari}}, \bibinfo {author} {\bibfnamefont {G.}~\bibnamefont {Harder}},
  \bibinfo {author} {\bibfnamefont {M.}~\bibnamefont {Allgaier}}, \bibinfo
  {author} {\bibfnamefont {B.}~\bibnamefont {Brecht}}, \ and\ \bibinfo {author}
  {\bibfnamefont {C.}~\bibnamefont {Silberhorn}},\ } \newblock \bibinfo{title}{Temporal-mode measurement tomography of a quantum pulse gate,} \href{https://journals.aps.org/pra/abstract/10.1103/PhysRevA.96.063817}{\bibfield
  {journal} {\bibinfo  {journal} {Phys. Rev. A}\ }\textbf {\bibinfo
  {volume} {96}},\ \bibinfo {pages} {063817} (\bibinfo {year}
  {2017})}\BibitemShut {NoStop}%
\bibitem [{\citenamefont {Liao}\ \emph
  {et~al.}(2017{\natexlab{a}})\citenamefont {Liao}, \citenamefont {Cai},
  \citenamefont {Liu}, \citenamefont {Zhang}, \citenamefont {Li}, \citenamefont
  {Ren}, \citenamefont {Yin}, \citenamefont {Shen}, \citenamefont {Cao},
  \citenamefont {Li} \emph {et~al.}}]{liao2017satellite}%
  \BibitemOpen
  \bibfield  {author} {\bibinfo {author} {\bibfnamefont {S.-K.}\ \bibnamefont
  {Liao}}, \bibinfo {author} {\bibfnamefont {W.-Q.}\ \bibnamefont {Cai}},
  \bibinfo {author} {\bibfnamefont {W.-Y.}\ \bibnamefont {Liu}}, \bibinfo
  {author} {\bibfnamefont {L.}~\bibnamefont {Zhang}}, \bibinfo {author}
  {\bibfnamefont {Y.}~\bibnamefont {Li}}, \bibinfo {author} {\bibfnamefont
  {J.-G.}\ \bibnamefont {Ren}}, \bibinfo {author} {\bibfnamefont
  {J.}~\bibnamefont {Yin}}, \bibinfo {author} {\bibfnamefont {Q.}~\bibnamefont
  {Shen}}, \bibinfo {author} {\bibfnamefont {Y.}~\bibnamefont {Cao}}, \bibinfo
  {author} {\bibfnamefont {Z.-P.}\ \bibnamefont {Li}},  \emph {et~al.},\
  }\newblock \bibinfo{title}{Satellite-to-ground quantum key distribution,} \href{https://www.nature.com/articles/nature23655/}{\href@noop {} {\bibfield  {journal} {\bibinfo  {journal} {Nature}\ }\textbf
  {\bibinfo {volume} {549}},\ \bibinfo {pages} {43} (\bibinfo {year}
  {2017}{\natexlab{a}})}}\BibitemShut {NoStop}%
\bibitem [{\citenamefont {Liu}\ \emph {et~al.}(2010)\citenamefont {Liu},
  \citenamefont {Chen}, \citenamefont {Wang}, \citenamefont {Cai},
  \citenamefont {Wan}, \citenamefont {Chen}, \citenamefont {Wang},
  \citenamefont {Liu}, \citenamefont {Liang}, \citenamefont {Yang} \emph
  {et~al.}}]{liu2010decoy}%
  \BibitemOpen
  \bibfield  {author} {\bibinfo {author} {\bibfnamefont {Y.}~\bibnamefont
  {Liu}}, \bibinfo {author} {\bibfnamefont {T.-Y.}\ \bibnamefont {Chen}},
  \bibinfo {author} {\bibfnamefont {J.}~\bibnamefont {Wang}}, \bibinfo {author}
  {\bibfnamefont {W.-Q.}\ \bibnamefont {Cai}}, \bibinfo {author} {\bibfnamefont
  {X.}~\bibnamefont {Wan}}, \bibinfo {author} {\bibfnamefont {L.-K.}\
  \bibnamefont {Chen}}, \bibinfo {author} {\bibfnamefont {J.-H.}\ \bibnamefont
  {Wang}}, \bibinfo {author} {\bibfnamefont {S.-B.}\ \bibnamefont {Liu}},
  \bibinfo {author} {\bibfnamefont {H.}~\bibnamefont {Liang}}, \bibinfo
  {author} {\bibfnamefont {L.}~\bibnamefont {Yang}}, \bibinfo
  {author} {\bibfnamefont {C.-Z.}~\bibnamefont {Peng}}, \bibinfo
  {author} {\bibfnamefont {K.}~\bibnamefont {Chen}}, \bibinfo
  {author} {\bibfnamefont {Z.-B.}~\bibnamefont {Chen}}, and \bibinfo
  {author} {\bibfnamefont {J.-W.}~\bibnamefont {Pan}},\
  }\newblock \bibinfo{title}{Decoy-state quantum key distribution with polarized photons over 200 km,} \href{https://doi.org/10.1364/OE.18.008587} {\bibfield  {journal} {\bibinfo  {journal} {Opt. Exp.}\
  }\textbf {\bibinfo {volume} {18}},\ \bibinfo {pages} {8587} (\bibinfo {year}
  {2010})}\BibitemShut {NoStop}%
\bibitem [{\citenamefont {Boaron}\ \emph {et~al.}(2018)\citenamefont {Boaron},
  \citenamefont {Boso}, \citenamefont {Rusca}, \citenamefont {Vulliez},
  \citenamefont {Autebert}, \citenamefont {Caloz}, \citenamefont {Perrenoud},
  \citenamefont {Gras}, \citenamefont {Bussi{\`e}res}, \citenamefont {Li} \emph
  {et~al.}}]{boaron2018secure}%
  \BibitemOpen
  \bibfield  {author} {\bibinfo {author} {\bibfnamefont {A.}~\bibnamefont
  {Boaron}}, \bibinfo {author} {\bibfnamefont {G.}~\bibnamefont {Boso}},
  \bibinfo {author} {\bibfnamefont {D.}~\bibnamefont {Rusca}}, \bibinfo
  {author} {\bibfnamefont {C.}~\bibnamefont {Vulliez}}, \bibinfo {author}
  {\bibfnamefont {C.}~\bibnamefont {Autebert}}, \bibinfo {author}
  {\bibfnamefont {M.}~\bibnamefont {Caloz}}, \bibinfo {author} {\bibfnamefont
  {M.}~\bibnamefont {Perrenoud}}, \bibinfo {author} {\bibfnamefont
  {G.}~\bibnamefont {Gras}}, \bibinfo {author} {\bibfnamefont {F.}~\bibnamefont
  {Bussi{\`e}res}}, \bibinfo {author} {\bibfnamefont {M.-J.}\ \bibnamefont
  {Li}}, \bibinfo {author} {\bibfnamefont {D.}~\bibnamefont
  {Nolan}}, \bibinfo {author} {\bibfnamefont {A.}~\bibnamefont
  {Martin}}, and \bibinfo {author} {\bibfnamefont {H.}~\bibnamefont
  {Zbinden}},\ }\newblock \bibinfo{title}{Secure Quantum Key Distribution over 421 km of Optical Fiber,} \href{https://journals.aps.org/prl/abstract/10.1103/PhysRevLett.121.190502} {\bibfield  {journal} {\bibinfo
  {journal} {Phys. Rev. Lett.}\ }\textbf {\bibinfo {volume} {121}},\
  \bibinfo {pages} {190502} (\bibinfo {year} {2018})}\BibitemShut {NoStop}%
\bibitem [{\citenamefont {Jin}\ \emph {et~al.}(2019)\citenamefont {Jin},
  \citenamefont {Bourgoin}, \citenamefont {Tannous}, \citenamefont {Agne},
  \citenamefont {Pugh}, \citenamefont {Kuntz}, \citenamefont {Higgins},\ and\
  \citenamefont {Jennewein}}]{jin2019genuine}%
  \BibitemOpen
  \bibfield  {author} {\bibinfo {author} {\bibfnamefont {J.}~\bibnamefont
  {Jin}}, \bibinfo {author} {\bibfnamefont {J.-P.}\ \bibnamefont {Bourgoin}},
  \bibinfo {author} {\bibfnamefont {R.}~\bibnamefont {Tannous}}, \bibinfo
  {author} {\bibfnamefont {S.}~\bibnamefont {Agne}}, \bibinfo {author}
  {\bibfnamefont {C.~J.}\ \bibnamefont {Pugh}}, \bibinfo {author}
  {\bibfnamefont {K.~B.}\ \bibnamefont {Kuntz}}, \bibinfo {author}
  {\bibfnamefont {B.~L.}\ \bibnamefont {Higgins}}, \ and\ \bibinfo {author}
  {\bibfnamefont {T.}~\bibnamefont {Jennewein}},\ }\newblock \bibinfo{title}{Genuine time-bin-encoded quantum key distribution over a turbulent depolarizing free-space channel,} \href{https://www.osapublishing.org/oe/fulltext.cfm?uri=oe-27-26-37214&id=423951} {\bibfield
  {journal} {\bibinfo  {journal} {Opt. Exp.}\ }\textbf {\bibinfo {volume}
  {27}},\ \bibinfo {pages} {37214} (\bibinfo {year} {2019})}\BibitemShut
  {NoStop}%
\bibitem [{\citenamefont {Sasaki}\ \emph {et~al.}(2014)\citenamefont {Sasaki},
  \citenamefont {Yamamoto},\ and\ \citenamefont
  {Koashi}}]{sasaki2014practical}%
  \BibitemOpen
  \bibfield  {author} {\bibinfo {author} {\bibfnamefont {T.}~\bibnamefont
  {Sasaki}}, \bibinfo {author} {\bibfnamefont {Y.}~\bibnamefont {Yamamoto}}, \
  and\ \bibinfo {author} {\bibfnamefont {M.}~\bibnamefont {Koashi}},\
  }\newblock \bibinfo{title}{Practical quantum key distribution protocol without monitoring signal disturbance,} \href{https://www.nature.com/articles/nature13303} {\bibfield  {journal} {\bibinfo  {journal} {Nature}\ }\textbf
  {\bibinfo {volume} {509}},\ \bibinfo {pages} {475} (\bibinfo {year}
  {2014})}\BibitemShut {NoStop}%
\bibitem [{\citenamefont {Ecker}\ \emph {et~al.}(2019)\citenamefont {Ecker},
  \citenamefont {Bouchard}, \citenamefont {Bulla}, \citenamefont {Brandt},
  \citenamefont {Kohout}, \citenamefont {Steinlechner}, \citenamefont
  {Fickler}, \citenamefont {Malik}, \citenamefont {Guryanova}, \citenamefont
  {Ursin},\ and\ \citenamefont {Huber}}]{ecker2019overcoming}%
  \BibitemOpen
  \bibfield  {author} {\bibinfo {author} {\bibfnamefont {S.}~\bibnamefont
  {Ecker}}, \bibinfo {author} {\bibfnamefont {F.}~\bibnamefont {Bouchard}},
  \bibinfo {author} {\bibfnamefont {L.}~\bibnamefont {Bulla}}, \bibinfo
  {author} {\bibfnamefont {F.}~\bibnamefont {Brandt}}, \bibinfo {author}
  {\bibfnamefont {O.}~\bibnamefont {Kohout}}, \bibinfo {author} {\bibfnamefont
  {F.}~\bibnamefont {Steinlechner}}, \bibinfo {author} {\bibfnamefont
  {R.}~\bibnamefont {Fickler}}, \bibinfo {author} {\bibfnamefont
  {M.}~\bibnamefont {Malik}}, \bibinfo {author} {\bibfnamefont
  {Y.}~\bibnamefont {Guryanova}}, \bibinfo {author} {\bibfnamefont
  {R.}~\bibnamefont {Ursin}}, \ and\ \bibinfo {author} {\bibfnamefont
  {M.}~\bibnamefont {Huber}},\ }\newblock \bibinfo{title}{Overcoming Noise in Entanglement Distribution,} \href {\doibase 10.1103/PhysRevX.9.041042}
  {\bibfield  {journal} {\bibinfo  {journal} {Phys. Rev. X}\ }\textbf {\bibinfo
  {volume} {9}},\ \bibinfo {pages} {041042} (\bibinfo {year}
  {2019})}\BibitemShut {NoStop}%
\bibitem [{\citenamefont {Marcikic}\ \emph {et~al.}(2002)\citenamefont
  {Marcikic}, \citenamefont {de~Riedmatten}, \citenamefont {Tittel},
  \citenamefont {Scarani}, \citenamefont {Zbinden},\ and\ \citenamefont
  {Gisin}}]{marcikic2002time}%
  \BibitemOpen
  \bibfield  {author} {\bibinfo {author} {\bibfnamefont {I.}~\bibnamefont
  {Marcikic}}, \bibinfo {author} {\bibfnamefont {H.}~\bibnamefont
  {de~Riedmatten}}, \bibinfo {author} {\bibfnamefont {W.}~\bibnamefont
  {Tittel}}, \bibinfo {author} {\bibfnamefont {V.}~\bibnamefont {Scarani}},
  \bibinfo {author} {\bibfnamefont {H.}~\bibnamefont {Zbinden}}, \ and\
  \bibinfo {author} {\bibfnamefont {N.}~\bibnamefont {Gisin}},\ }\newblock \bibinfo{title}{Time-bin entangled qubits for quantum communication created by femtosecond pulses,} \href{https://journals.aps.org/pra/abstract/10.1103/PhysRevA.66.062308}
  {\bibfield  {journal} {\bibinfo  {journal} {Phys. Rev. A}\ }\textbf
  {\bibinfo {volume} {66}},\ \bibinfo {pages} {062308} (\bibinfo {year}
  {2002})}\BibitemShut {NoStop}%
\bibitem [{\citenamefont {Brougham}\ \emph {et~al.}(2013)\citenamefont
  {Brougham}, \citenamefont {Barnett}, \citenamefont {McCusker}, \citenamefont
  {Kwiat},\ and\ \citenamefont {Gauthier}}]{brougham2013security}%
  \BibitemOpen
  \bibfield  {author} {\bibinfo {author} {\bibfnamefont {T.}~\bibnamefont
  {Brougham}}, \bibinfo {author} {\bibfnamefont {S.~M.}\ \bibnamefont
  {Barnett}}, \bibinfo {author} {\bibfnamefont {K.~T.}\ \bibnamefont
  {McCusker}}, \bibinfo {author} {\bibfnamefont {P.~G.}\ \bibnamefont {Kwiat}},
  \ and\ \bibinfo {author} {\bibfnamefont {D.~J.}\ \bibnamefont {Gauthier}},\
  }\newblock \bibinfo{title}{Security of high-dimensional quantum key distribution protocols using Franson interferometers,} \href {https://iopscience.iop.org/article/10.1088/0953-4075/46/10/104010} {\bibfield  {journal} {\bibinfo  {journal} {J. Phys. B}\ }\textbf {\bibinfo {volume}
  {46}},\ \bibinfo {pages} {104010} (\bibinfo {year} {2013})}\BibitemShut
  {NoStop}%
\bibitem [{\citenamefont {Caloz}\ \emph {et~al.}(2018)\citenamefont {Caloz},
  \citenamefont {Perrenoud}, \citenamefont {Autebert}, \citenamefont {Korzh},
  \citenamefont {Weiss}, \citenamefont {Sch{\"o}nenberger}, \citenamefont
  {Warburton}, \citenamefont {Zbinden},\ and\ \citenamefont
  {Bussi{\`e}res}}]{caloz2018high}%
  \BibitemOpen
  \bibfield  {author} {\bibinfo {author} {\bibfnamefont {M.}~\bibnamefont
  {Caloz}}, \bibinfo {author} {\bibfnamefont {M.}~\bibnamefont {Perrenoud}},
  \bibinfo {author} {\bibfnamefont {C.}~\bibnamefont {Autebert}}, \bibinfo
  {author} {\bibfnamefont {B.}~\bibnamefont {Korzh}}, \bibinfo {author}
  {\bibfnamefont {M.}~\bibnamefont {Weiss}}, \bibinfo {author} {\bibfnamefont
  {C.}~\bibnamefont {Sch{\"o}nenberger}}, \bibinfo {author} {\bibfnamefont
  {R.~J.}\ \bibnamefont {Warburton}}, \bibinfo {author} {\bibfnamefont
  {H.}~\bibnamefont {Zbinden}}, \ and\ \bibinfo {author} {\bibfnamefont
  {F.}~\bibnamefont {Bussi{\`e}res}},\ }\newblock \bibinfo{title}{High-detection efficiency and low-timing jitter with amorphous superconducting nanowire single-photon detectors,} \href{https://aip.scitation.org/doi/abs/10.1063/1.5010102} {\bibfield  {journal}
  {\bibinfo  {journal} {Appl. Phys. Lett.}\ }\textbf {\bibinfo {volume}
  {112}},\ \bibinfo {pages} {061103} (\bibinfo {year} {2018})}\BibitemShut
  {NoStop}%
\bibitem [{\citenamefont {Korzh}\ \emph {et~al.}(2020)\citenamefont {Korzh},
  \citenamefont {Zhao}, \citenamefont {Allmaras}, \citenamefont {Frasca},
  \citenamefont {Autry}, \citenamefont {Bersin}, \citenamefont {Beyer},
  \citenamefont {Briggs}, \citenamefont {Bumble}, \citenamefont {Colangelo}
  \emph {et~al.}}]{korzh2020demonstration}%
  \BibitemOpen
  \bibfield  {author} {\bibinfo {author} {\bibfnamefont {B.}~\bibnamefont
  {Korzh}}, \bibinfo {author} {\bibfnamefont {Q.-Y.}\ \bibnamefont {Zhao}},
  \bibinfo {author} {\bibfnamefont {J.~P.}\ \bibnamefont {Allmaras}}, \bibinfo
  {author} {\bibfnamefont {S.}~\bibnamefont {Frasca}}, \bibinfo {author}
  {\bibfnamefont {T.~M.}\ \bibnamefont {Autry}}, \bibinfo {author}
  {\bibfnamefont {E.~A.}\ \bibnamefont {Bersin}}, \bibinfo {author}
  {\bibfnamefont {A.~D.}\ \bibnamefont {Beyer}}, \bibinfo {author}
  {\bibfnamefont {R.~M.}\ \bibnamefont {Briggs}}, \bibinfo {author}
  {\bibfnamefont {B.}~\bibnamefont {Bumble}}, \bibinfo {author} {\bibfnamefont
  {M.}~\bibnamefont {Colangelo}},  \emph {et~al.},\ }\newblock \bibinfo{title}{Demonstration of sub-3 ps temporal resolution with a superconducting nanowire single-photon detector,} \href{https://www.nature.com/articles/s41566-020-0589-x} {\bibfield
  {journal} {\bibinfo  {journal} {Nat. Photonics}\ }\textbf {\bibinfo
  {volume} {14}},\ \bibinfo {pages} {250} (\bibinfo {year} {2020})}\BibitemShut
  {NoStop}%
\bibitem [{\citenamefont {Islam}\ \emph
  {et~al.}(2017{\natexlab{a}})\citenamefont {Islam}, \citenamefont {Lim},
  \citenamefont {Cahall}, \citenamefont {Kim},\ and\ \citenamefont
  {Gauthier}}]{islam2017provably}%
  \BibitemOpen
  \bibfield  {author} {\bibinfo {author} {\bibfnamefont {N.~T.}\ \bibnamefont
  {Islam}}, \bibinfo {author} {\bibfnamefont {C.~C.~W.}\ \bibnamefont {Lim}},
  \bibinfo {author} {\bibfnamefont {C.}~\bibnamefont {Cahall}}, \bibinfo
  {author} {\bibfnamefont {J.}~\bibnamefont {Kim}}, \ and\ \bibinfo {author}
  {\bibfnamefont {D.~J.}\ \bibnamefont {Gauthier}},\ }\newblock \bibinfo{title}{Provably secure and high-rate quantum key distribution with time-bin qudits,} \href{https://advances.sciencemag.org/content/3/11/e1701491} {\bibfield
  {journal} {\bibinfo  {journal} {Sci. Adv.}\ }\textbf {\bibinfo
  {volume} {3}},\ \bibinfo {pages} {e1701491} (\bibinfo {year}
  {2017}{\natexlab{a}})}\BibitemShut {NoStop}%
\bibitem [{\citenamefont {Islam}\ \emph
  {et~al.}(2017{\natexlab{b}})\citenamefont {Islam}, \citenamefont {Cahall},
  \citenamefont {Aragoneses}, \citenamefont {Lezama}, \citenamefont {Kim},\
  and\ \citenamefont {Gauthier}}]{islam2017robust}%
  \BibitemOpen
  \bibfield  {author} {\bibinfo {author} {\bibfnamefont {N.~T.}\ \bibnamefont
  {Islam}}, \bibinfo {author} {\bibfnamefont {C.}~\bibnamefont {Cahall}},
  \bibinfo {author} {\bibfnamefont {A.}~\bibnamefont {Aragoneses}}, \bibinfo
  {author} {\bibfnamefont {A.}~\bibnamefont {Lezama}}, \bibinfo {author}
  {\bibfnamefont {J.}~\bibnamefont {Kim}}, \ and\ \bibinfo {author}
  {\bibfnamefont {D.~J.}\ \bibnamefont {Gauthier}},\ }\newblock \bibinfo{title}{Robust and Stable Delay Interferometers with Application to 
\textit{d}-Dimensional Time-Frequency Quantum Key Distribution,} \href{https://journals.aps.org/prapplied/abstract/10.1103/PhysRevApplied.7.044010} {\bibfield
  {journal} {\bibinfo  {journal} {Phys. Rev. Appl.}\ }\textbf {\bibinfo
  {volume} {7}},\ \bibinfo {pages} {044010} (\bibinfo {year}
  {2017}{\natexlab{b}})}\BibitemShut {NoStop}%
\bibitem [{\citenamefont {MacLean}\ \emph {et~al.}(2018)\citenamefont
  {MacLean}, \citenamefont {Donohue},\ and\ \citenamefont
  {Resch}}]{maclean2018direct}%
  \BibitemOpen
  \bibfield  {author} {\bibinfo {author} {\bibfnamefont {J.-P.~W.}\
  \bibnamefont {MacLean}}, \bibinfo {author} {\bibfnamefont {J.~M.}\
  \bibnamefont {Donohue}}, \ and\ \bibinfo {author} {\bibfnamefont {K.~J.}\
  \bibnamefont {Resch}},\ }\newblock \bibinfo{title}{Direct Characterization of Ultrafast Energy-Time Entangled Photon Pairs,} \href{https://journals.aps.org/prl/abstract/10.1103/PhysRevLett.120.053601} {\bibfield  {journal} {\bibinfo
  {journal} {Phys. Rev. Lett.}\ }\textbf {\bibinfo {volume} {120}},\
  \bibinfo {pages} {053601} (\bibinfo {year} {2018})}\BibitemShut {NoStop}%
\bibitem [{\citenamefont {England}\ \emph {et~al.}(2015)\citenamefont
  {England}, \citenamefont {Fisher}, \citenamefont {MacLean}, \citenamefont
  {Bustard}, \citenamefont {Lausten}, \citenamefont {Resch},\ and\
  \citenamefont {Sussman}}]{england2015storage}%
  \BibitemOpen
  \bibfield  {author} {\bibinfo {author} {\bibfnamefont {D.~G.}\ \bibnamefont
  {England}}, \bibinfo {author} {\bibfnamefont {K.~A.}\ \bibnamefont {Fisher}},
  \bibinfo {author} {\bibfnamefont {J.-P.~W.}\ \bibnamefont {MacLean}},
  \bibinfo {author} {\bibfnamefont {P.~J.}\ \bibnamefont {Bustard}}, \bibinfo
  {author} {\bibfnamefont {R.}~\bibnamefont {Lausten}}, \bibinfo {author}
  {\bibfnamefont {K.~J.}\ \bibnamefont {Resch}}, \ and\ \bibinfo {author}
  {\bibfnamefont {B.~J.}\ \bibnamefont {Sussman}},\ }\newblock \bibinfo{title}{Storage and Retrieval of THz-Bandwidth Single Photons Using a Room-Temperature Diamond Quantum Memory,} \href{https://journals.aps.org/prl/abstract/10.1103/PhysRevLett.114.053602} {\bibfield
  {journal} {\bibinfo  {journal} {Phys. Rev. Lett.}\ }\textbf {\bibinfo
  {volume} {114}},\ \bibinfo {pages} {053602} (\bibinfo {year}
  {2015})}\BibitemShut {NoStop}%
\bibitem [{\citenamefont {Kupchak}\ \emph {et~al.}(2017)\citenamefont
  {Kupchak}, \citenamefont {Bustard}, \citenamefont {Heshami}, \citenamefont
  {Erskine}, \citenamefont {Spanner}, \citenamefont {England},\ and\
  \citenamefont {Sussman}}]{kupchak2017time}%
  \BibitemOpen
  \bibfield  {author} {\bibinfo {author} {\bibfnamefont {C.}~\bibnamefont
  {Kupchak}}, \bibinfo {author} {\bibfnamefont {P.~J.}\ \bibnamefont
  {Bustard}}, \bibinfo {author} {\bibfnamefont {K.}~\bibnamefont {Heshami}},
  \bibinfo {author} {\bibfnamefont {J.}~\bibnamefont {Erskine}}, \bibinfo
  {author} {\bibfnamefont {M.}~\bibnamefont {Spanner}}, \bibinfo {author}
  {\bibfnamefont {D.~G.}\ \bibnamefont {England}}, \ and\ \bibinfo {author}
  {\bibfnamefont {B.~J.}\ \bibnamefont {Sussman}},\ }\newblock \bibinfo{title}{Time-bin-to-polarization conversion of ultrafast photonic qubits,} \href{https://journals.aps.org/pra/abstract/10.1103/PhysRevA.96.053812} {\bibfield
  {journal} {\bibinfo  {journal} {Phys. Rev. A}\ }\textbf {\bibinfo
  {volume} {96}},\ \bibinfo {pages} {053812} (\bibinfo {year}
  {2017})}\BibitemShut {NoStop}%
\bibitem [{\citenamefont {Kupchak}\ \emph {et~al.}(2019)\citenamefont
  {Kupchak}, \citenamefont {Erskine}, \citenamefont {England},\ and\
  \citenamefont {Sussman}}]{kupchak2019terahertz}%
  \BibitemOpen
  \bibfield  {author} {\bibinfo {author} {\bibfnamefont {C.}~\bibnamefont
  {Kupchak}}, \bibinfo {author} {\bibfnamefont {J.}~\bibnamefont {Erskine}},
  \bibinfo {author} {\bibfnamefont {D.}~\bibnamefont {England}}, \ and\
  \bibinfo {author} {\bibfnamefont {B.}~\bibnamefont {Sussman}},\ }\newblock \bibinfo{title}{Terahertz-bandwidth switching of heralded single photons,} \href{https://www.osapublishing.org/ol/abstract.cfm?uri=ol-44-6-1427} {\bibfield  {journal} {\bibinfo  {journal} {Opt. Lett.}\ }\textbf
  {\bibinfo {volume} {44}},\ \bibinfo {pages} {1427} (\bibinfo {year}
  {2019})}\BibitemShut {NoStop}%
\bibitem [{\citenamefont {Ma}\ \emph {et~al.}(2005)\citenamefont {Ma},
  \citenamefont {Qi}, \citenamefont {Zhao},\ and\ \citenamefont
  {Lo}}]{ma2005practical}%
  \BibitemOpen
  \bibfield  {author} {\bibinfo {author} {\bibfnamefont {X.}~\bibnamefont
  {Ma}}, \bibinfo {author} {\bibfnamefont {B.}~\bibnamefont {Qi}}, \bibinfo
  {author} {\bibfnamefont {Y.}~\bibnamefont {Zhao}}, \ and\ \bibinfo {author}
  {\bibfnamefont {H.-K.}\ \bibnamefont {Lo}},\ }\newblock \bibinfo{title}{Practical decoy state for quantum key distribution,} \href{https://journals.aps.org/pra/abstract/10.1103/PhysRevA.72.012326} {\bibfield
  {journal} {\bibinfo  {journal} {Phys. Rev. A}\ }\textbf {\bibinfo
  {volume} {72}},\ \bibinfo {pages} {012326} (\bibinfo {year}
  {2005})}\BibitemShut {NoStop}%
\bibitem [{\citenamefont {Stucki}\ \emph {et~al.}(2005)\citenamefont {Stucki},
  \citenamefont {Brunner}, \citenamefont {Gisin}, \citenamefont {Scarani},\
  and\ \citenamefont {Zbinden}}]{stucki2005fast}%
  \BibitemOpen
  \bibfield  {author} {\bibinfo {author} {\bibfnamefont {D.}~\bibnamefont
  {Stucki}}, \bibinfo {author} {\bibfnamefont {N.}~\bibnamefont {Brunner}},
  \bibinfo {author} {\bibfnamefont {N.}~\bibnamefont {Gisin}}, \bibinfo
  {author} {\bibfnamefont {V.}~\bibnamefont {Scarani}}, \ and\ \bibinfo
  {author} {\bibfnamefont {H.}~\bibnamefont {Zbinden}},\ }\newblock \bibinfo{title}{Fast and simple one-way quantum key distribution,} \href{https://aip.scitation.org/doi/10.1063/1.2126792}
  {\bibfield  {journal} {\bibinfo  {journal} {Appl. Phys. Lett.}\
  }\textbf {\bibinfo {volume} {87}},\ \bibinfo {pages} {194108} (\bibinfo
  {year} {2005})}\BibitemShut {NoStop}%
\bibitem [{\citenamefont {Bru{\ss}}(1998)}]{bruss1998optimal}%
  \BibitemOpen
  \bibfield  {author} {\bibinfo {author} {\bibfnamefont {D.}~\bibnamefont
  {Bru{\ss}}},\ }\newblock \bibinfo{title}{Optimal Eavesdropping in Quantum Cryptography with Six States,} \href{https://journals.aps.org/prl/abstract/10.1103/PhysRevLett.81.3018} {\bibfield  {journal} {\bibinfo  {journal}
  {Phys. Rev. Lett.}\ }\textbf {\bibinfo {volume} {81}},\ \bibinfo
  {pages} {3018} (\bibinfo {year} {1998})}\BibitemShut {NoStop}%
\bibitem [{\citenamefont {Bouchard}\ \emph {et~al.}(2021)\citenamefont
  {Bouchard}, \citenamefont {England}, \citenamefont {Bustard}, \citenamefont
  {Fenwick}, \citenamefont {Karimi}, \citenamefont {Heshami},\ and\
  \citenamefont {Sussman}}]{bouchard2021achieving}%
  \BibitemOpen
  \bibfield  {author} {\bibinfo {author} {\bibfnamefont {F.}~\bibnamefont
  {Bouchard}}, \bibinfo {author} {\bibfnamefont {D.}~\bibnamefont {England}},
  \bibinfo {author} {\bibfnamefont {P.~J.}\ \bibnamefont {Bustard}}, \bibinfo
  {author} {\bibfnamefont {K.~L.}\ \bibnamefont {Fenwick}}, \bibinfo {author}
  {\bibfnamefont {E.}~\bibnamefont {Karimi}}, \bibinfo {author} {\bibfnamefont
  {K.}~\bibnamefont {Heshami}}, \ and\ \bibinfo {author} {\bibfnamefont
  {B.}~\bibnamefont {Sussman}},\ }\newblock \bibinfo{title}{Achieving Ultimate Noise Tolerance in Quantum Communication,} \href{https://journals.aps.org/prapplied/abstract/10.1103/PhysRevApplied.15.024027} {\bibfield  {journal} {\bibinfo
   {journal} {Phys. Rev. Appl.}\ }\textbf {\bibinfo {volume} {15}},\
  \bibinfo {pages} {024027} (\bibinfo {year} {2021})}\BibitemShut {NoStop}%
\bibitem [{\citenamefont {Raymer}\ and\ \citenamefont
  {Banaszek}(2020)}]{raymer2020time}%
  \BibitemOpen
  \bibfield  {author} {\bibinfo {author} {\bibfnamefont {M.~G.}\ \bibnamefont
  {Raymer}}\ and\ \bibinfo {author} {\bibfnamefont {K.}~\bibnamefont
  {Banaszek}},\ }\newblock \bibinfo{title}{Time-frequency optical filtering: efficiency vs. temporal-mode discrimination in incoherent and coherent implementations,} \href{https://www.osapublishing.org/oe/fulltext.cfm?uri=oe-28-22-32819&id=441685} {\bibfield  {journal} {\bibinfo  {journal}
  {Opt. Exp.}\ }\textbf {\bibinfo {volume} {28}},\ \bibinfo {pages}
  {32819} (\bibinfo {year} {2020})}\BibitemShut {NoStop}%
\end{thebibliography}

%

\end{document}